\documentclass[prb, twocolumn, showpacs, amsmath, amssymb]{revtex4}
\usepackage{dcolumn}
\usepackage{bm}
\usepackage{graphicx}
\usepackage{color}
\usepackage{bm}
\begin{document}
\pacs{ 75.50.-y, 76.60.Es, 75.60.Gb}
\title{Observation of different spin behavior with temperature variation and Cr substitution in a multiferroic compound YMn$_2$O$_5$}

\author{K. Mukherjee}
\altaffiliation[Present address: ]{Department of Materials Engineering Science, Graduate School of Engineering Science, Osaka University, Osaka 560-8531, Japan.}
\affiliation{UGC-DAE Consortium for Scientific Research\\University Campus, Khandwa Road\\
Indore-45201, M.P, India.}
\author{Kranti Kumar}
\affiliation{UGC-DAE Consortium for Scientific Research\\University Campus, Khandwa Road\\
Indore-45201, M.P, India.}
\author{A. Banerjee}
\affiliation{UGC-DAE Consortium for Scientific Research\\University Campus, Khandwa Road\\
Indore-45201, M.P, India.}

\begin{abstract}
In this article, the collective response of the spins is explored through low field bulk magnetic measurement for the series YMn$_{2-x}$Cr$_x$O$_5$ (x= 0.0, 0.05). Low field ac susceptibility and dc magnetization of YMn$_2$O$_5$ shows multiple transition in analogy to those observed in electrical measurement of the compound. Using various time dependent magnetization protocols it has been observed that the behavior of spins in commensurate and incommensurate phase are drastically different. YMn$_{1.95}$Cr$_{0.05}$O$_5$ undergoes a ferrimagnetic ordering with an enhanced magnetic ordering temperature as compared to the parent, which undergoes an antiferromagnetic ordering. Appearance of spontaneous magnetization without any major change in the atomic structure is rather significant since the parent compound is an important multiferroic material. In addition, magnetic memory effect is observed in the Cr substituted compound whereas it is absent in the parent compound.
\end{abstract}

\maketitle
\section {Introduction}
In the field of condensed matter physics study of materials exhibiting functional properties have emerged as one of the topic of great current interest. In this context, multiferroics (a compound that exhibits multiple long-range ordering among magnetic, electric and/or elastic properties) have been the subject of extensive research in recent years, both from the viewpoint of technological applications and the underlying physics.\cite{lee,kim,hur,cha,bla,cha1, mam, nod, yam, sus, woo, fab} In these materials electric polarization can be switched by magnetic field and magnetization by an electric field, a property useful for application in magnetic storage media and spintronics. Also the interesting physics associated with the mutual coupling between electric, magnetic and lattice degrees of freedom is worth probing. In these context it should be noted that there exists different mechanism of ferroelectric ordering in different types of ferroelectrics\cite{kho}, but the origin of magnetic ordering is same for all cases and is due to an exchange interaction of predominantly localized magnetic moments. However due to contrasting requirements (i.e. empty d orbitals for ferroelectricity and partially filled d orbitals for magnetic ordering) materials exhibiting multiferroicity are not so common.

In systems known as improper multiferroics, magnetic ordering induces ferroelectric ordering, giving rise to the possibility of strong coupling of ferroelectricity with magnetic ordering. In this family, the ferroelectric transition temperature lies below the magnetic ordering temperature, as generally observed in RMn$_2$O$_5$ (R = rare earth ions) series.\cite{ino, hur1, kim1, sou, par, hiro, kim3} Here phenomena like incommensurate  magnetic ordering along with spin frustration, lock in transition and ferroelectricity is observed within a short temperature interval. Generally all these materials on cooling below antiferromagnetic transition temperature, undergo multiple phase transitions involving changes in magnetic propagation vector k = (k$_x$, k$_y$, k$_z$) and the changes often coincides with pronounced anomalies in temperature response of dielectric constant, polarization, magnetization and specific heat.

In this family, YMn$_2$O$_5$ have emerged as an interesting candidate to study the multiferroics properties as Y$^{3+}$ ion is non-magnetic and the observed ordering is only due to Mn moments. In this system the study of spin behavior as temperature is varied is simplified due to absence of any interference from the rare earth site. For this purpose probes like low magnetic field and Mn site substitution can be useful tools. Even though extensive studies on YMn$_2$O$_5$ have been carried out using bulk as well as microscopic probes \cite{cha, sus, nod, sou, par, kim3}, investigation using above mentioned route are lacking in literature. Low magnetic field measurement are useful to distinguish the magnetic state of an inhomogeneously magnetized systems, while large applied fields can mask the intrinsic signatures. In addition, in colossal magnetoresistance manganites there are reports that substitution of Cr at Mn site can lead to a supression of antiferromagnetism in the parent compound and make the system ferromagnetic.\cite{kim2} But this route of investigation is rarely explored \cite{zhou} in this family where there can be a possibility to have a compound where there is a coexistance between ferromagnetism and ferrroelectricity. 

Hence, in this paper, through bulk magnetic measurement we have investigated the magnetic properties of the series YMn$_{2-x}$Cr$_x$O$_5$ (x = 0.0, 0.05). The main observation of this study are (i) signature of transitions observed from electrical measurements (which are extensive in literature) are reflected in low magnetic field studies of ac susceptibility and dc magnetization, ii) magnetic relaxation measurements brings out the differences in the spin behavior in the commensurate (CM) and incommensurate (ICM) phases, which coexists within a small temperature interval, and iii) introduction of Cr changes the magnetic ordering from antiferromagnetic to ferrimagnetic in the doped compound with an enhanced magnetic ordering temperature. Moreover, this substitution also bring in intriguing features like magnetic memory effect in the system.         
  
\maketitle\section{Sample preparation and characterization}
 
Polycrystalline YMn$_2$O$_5$ and YMn$_{1.95}$Cr$_{0.05}$O$_5$ samples are prepared by a chemical route called pyrophoric method.\cite{dey} Starting materials are aqueous solutions of stoichiometric quantities of high purity ($>$99.9\%) Y$_2$O$_3$, C$_4$H$_6$MnO$_4$.4H$_2$O and Cr(NO$_3$)$_3$. 9H$_2$O. The solutions are mixed together and triethanolamine (TEA) is added to it in such a way that the metal ion to TEA ratio is maintained at 1:2:2 (Y: Mn: TEA  =1: 2: 2). The viscous solution of TEA-complexed metal nitrates is evaporated on a hot plate with constant stirring. After complete dehydration, the nitrate themselves are decomposed with the evolution of brown fumes of NO$_2$ leaving behind organic-based, black, fluffy powders, which is the precursor powders. The dried carbon rich mass is grounded to fine powder and is calcined at 950$^0$C for 24hrs. The sample is then pelletized and heated in O$_2$ atmosphere at 1100$^0$C for 36hrs. X-Ray diffraction (XRD) was carried out using Rigaku Rotaflex RTC 300 RC diffractometer with CuK$\alpha$  radiation. All the samples are seen to be single phase and the pattern collected is analysed by the Rietveld profile refinement, using the profile refinement program by Young et al.\cite{you}  AC susceptibility is measured using a homemade susceptometer \cite{ashnarsi} and dc magnetization measurements in a commercial 14T vibrating sample magnetometer (M/s. Quantum Design, USA).  

All the samples crystallize in orthorhombic structure (Pbnm). Table 1 summarizes the relevant structural parameters obtained by fitting the powder XRD data by rietveld refinement. The Cr substitution does not introduce any significant structural distortion as is observed from lattice parameters summarized in Table 1.
\begin{table}[h]
\caption{\label{tab:table 1}Structural and fitting parameters determined from rietveld profile refinement of the XRD pattern of YMn$_{2-x}$Cr$_x$O$_5$ (x= 0.0, 0.05).}
\begin{ruledtabular}
\begin{tabular}{ccc}
Samples  &x = 0.0 & x = 0.05 \\
\hline
a(A${^O}$) &7.2603(1)   &7.2604(1) \\ 
\hline
b(A${^O}$) &8.4722(2)   &8.4748(2) \\ 
\hline
c(A${^O}$) &5.6649(1) &5.6675(1) \\
\hline
V(A${^O}$) &348.45   &348.72 \\
\hline
S &1.3 &1.28 \\
\end{tabular}
\end{ruledtabular}
\end{table} 

\maketitle\section{Results and discussions} 
\subsection{Low field magnetic studies of YMn$_2$O$_5$}
Temperature response of the real part of ac susceptibility ($\chi$$_1$$^R$) of the compound shows two peaks at 
$\approx$ 48K and 36K (Fig. 1a).  Zero field cooled magnetization (ZFCM) and field cooled magnetization (FCM) as a 
function of temperature measured in 200 Oe static magnetic field is shown in figure 1b. Analogous to the previous case, ZFCM curves shows two peaks, but at $\approx$ 42 and 34K. The differences in the peak temperatures between ac susceptibility and dc magnetization arises because of the differences in probing time scale of these two measurements 
and hence due to the consequent differences in the dynamics being probed. Also, it can be said that a higher value of dc magnetic field shift the peak down in temperature. Hence it can be said that the multiple phase transitions (involving changes in magnetic propagation vector) is sensitive to the applied magnetic field. (It is also to be noted that generally, the magnetic and ferroelectric transition temperature of different members of RMn$_2$O$_5$ family is found to vary between 39-45K and 32-39K respectively\cite{sat}). In our case, the peak at the higher temperature is associated with the ordering of Mn moments while that the lower one correspondence to lock-in temperatures associated with ICM-CM transition. The sequence of ordering observed in this compound is similar to isostructural TbMn$_2$O$_5$ where the magnetic structure is incommensurate [k=($\approx$0.50, 0, 0.30)] immediately below magnetic ordering temperature ($\approx$43K) and become commensurate with k=($\approx$0.50, 0, 0.25) on cooling through the lock in temperature at 33K.\cite{bla} Coming back to figure 1b, it is observed that in YMn$_2$O$_5$ FCM continuously rises as temperature is reduced, with the bifurcation between the ZFCM and FCM curves starting from 42K. Such a behavior of FCM curve indicates the absence of canonical spin glass or superparamagnetic behavior as for such cases FC curves saturates and shows a temperature independent behavior. This fact is further substantiated from frequency dependence of the real part of AC-$\chi$ (Fig 1c), which shows absence of shift in peak temperature with frequency indicating absence of metastable magnetic behavior. Temperature response of AC-$\chi$ at different fields (Fig 1d) demonstrates a lack of field dependence along with unchanged nature of the peaks indicating absence of magnetic clusters in the sample (and also indicating that the magnetic ordering is long ranged). It is to be noted that the presence of clusters results, in the change of shape of the peak with the variation of ac field. Hence from the above measurements, it can be said that the signatures of multiple phase transitions (involving changes in magnetic propagation vector) of YMn$_2$O$_5$ is also reflected in low field magnetic measurements.

\subsection{Investigation of spin behavior of YMn$_2$O$_5$ through time dependent magnetization measurement}
As stated before, the magnetic structure of the compound below the antiferromagnetic transition temperature is ICM whereas it becomes CM on cooling below $\approx$34K. On further cooling below $\approx$20K,  a new type of ICM phase reappears.\cite{sus} In the ICM case, each of the magnetic atoms in the unit cell has an independent spin density wave (SDW) i.e. its own amplitude and phase and there is no phase relation between the SDW's of different atoms. The phase contains variable mixtures of different configurations as it results from reversal of AFM zigzag chains along a-axis which give rise to four possible magnetic configuration per Mn$^{3+}$/Mn$^{4+}$ layer.\cite{bla} In CM case the phase on each crystallographic site within the chain is fixed in such a way that the moments follow a simple harmonic modulation. Hence it can be said that in the CM case all the atoms undergoes a similar type of alteration whereas in ICM case the modification among the neighboring atoms is not identical. Hence due to the presense of different types of spin configurations it would be very interesting to observe the nature of decay in magnetization through thermoremanent magnetization (TRM) measurement at different temperatures (three different temperature regions $\approx$ 48-32K, 32-20K and below 20K).  Each of the relaxation isotherms is obtained by cooling the sample in an applied field of 10kOe from 150K (the paramagnetic regime) to the measurement temperatures. The applied field is switched off at the temperature and magnetization decay is noted for the observation time of 3hrs. Among the various functional form that have been proposed to describe the change of magnetization with time, the relaxation at 35K and 25K is best represented by stretched exponential function given by
\begin{center}
M(t) = M$_0$ exp [-(t/$\tau$)$^n$].....(1)
\end{center}
while those at 15K and 2K is described quantitatively by the superposition of a stretched exponential and a constant:
\begin{center}
M(t) = M$_1$ + M$_0$ exp [-(t/$\tau$)$^n$].....(2)
\end{center}
In the above equations $\tau$ is the characteristic relaxation time, and M$_0$, M$_1$ and `n' are the fitting parameters. Fig. 2a shows the magnetization vs. time at different temperatures and the solid lines are the best fits of the experimental data. The fits yield a very good $\chi$$^2$ ($\approx$ 10$^{-6}$ - 10$^{-7}$) and small error bars in the fitting parameters. These values are compiled in Table 2. The initial magnetization (= M$_0$ + M$_1$) is seen to increase with decreasing temperature as is expected from the temperature dependence of FCM curve. From the value of $\tau$ it can be said that the relaxation time of spins in the ICM phase is more than that in the CM phase, while a faster relaxation is observed in the new ICM phase below 20K. Hence from the above measurement it can be said that the diverse transitions present in the system results in different spin configuration leading to different relaxation dynamics. The observed different types of relaxation dynamics seems to be consistent to those observed for reentrant compounds where the system enters an ordered ferromagnetic phase from a random paramagnetic phase and on further lowering of temperature re-enters once again into a new random phase. The relaxation behavior in reentrant phase is best described by a stretched exponential function and a constant.\cite{akm} In our case, the relaxation behavior at 15K and 2K (ICM phase) being best described by stretched exponential function and a constant is consistent to that observed for re-entrant systems.

\begin{table}
\caption{\label{tab:table 2} Fitting parameters obtained from equation (1) and (2)}
\begin{ruledtabular}
\begin{tabular}{ccccc}
T(K)  &35 &25 &15 &2 \\
\hline
M$_0$(emu/mol) &1.5767(8)   &7.3736(1) &0.3697(3) &0.4035(2)\\ 
\hline
M$_1$(emu/mol) &-   &- &10.2859(2) &14.1775(1)\\ 
\hline
n &0.150(4) &0.172(4) &0.479(6) &0.572(6)\\
\hline
$\tau$(sec) &1.068*10$^{13}$ &4.7*10$^{10}$ &2.0*10$^3$ &2.2*10$^3$\\
\hline
$\chi$$^2$ &10$^{-6}$ &10$^{-6}$ &10$^{-7}$ &10$^{-7}$ \\
\end{tabular}
\end{ruledtabular}
\end{table} 

To further investigate the difference in spin dynamics of ICM and CM phase, temperature dependent TRM measurements are done.\cite{muk} The sample is cooled from 200K to 38K in the ICM phase in zero field. At 38K the field (500Oe) is switched on and magnetization is noted for time t1 (1 hr). The sample was then warmed in same constant field to 42K and TRM is measure for time t2 (1 hr). Then the sample is cooled back to 38K in constant field and TRM is measured for time t3 (1hr). The same exercise is repeated in the CM phase with the measurement temperature being 22K, 26K and 22K. The curves are shown in figure 2b and 2c. In CM phase during t1 the curve shows an immediate rise followed by steady growth after the field is switched on. During temporary heating the magnetization value rises and a steady growth is observed at 26K. The relaxation curve at t3 is weak and the magnetization starts from a value it reached at the end of t2 as the FC values at 22 and 26K are almost similar. Hence the observed change in the magnetization values is in accordance to ZFCM and FCM curves. However the magnetization behavior in the ICM phase is quite different even though the measurement protocol is similar. In this phase the observed growth during t1 is less than that observed in the CM phase (i.e. at 22K). During t2 the relaxation is weak and magnetization value falls, which is not in accordance with the ZFCM curve. During t3 the magnetization rises to a value much higher than it reached at the end of t1 and a growth in magnetization with time is observed. Hence the above measurement in the two phases is quite contrasting, which brings out the differences in the spin arrangements in the CM and ICM phases, which coexists within a small temperature interval.  

Further, to get a better insight about the magnetic phases in the system, memory experiment under FC protocol is performed. In the experiment the sample is cooled from 300K in 100Oe with intermediary stops at the selected temperature of 45K, 35K and 25K for 2hours with the field switched off. At 45K, the magnetization becomes almost zero as the field is switched off. Moreover no decay is observed during the waiting time indicating the absence of relaxation effect and when the field is switched on magnetization value reaches same value it had before the waiting time.  At  the lower waiting temperatures aging effect is observed due to the presence of relaxation dynamics. After the waiting time at all temperatures, as the field is switched on it rejuvenates and returns to the zero age configuration. However on the warming cycle no memory of the wait temperatures is observed, which is a common feature for glassy systems. Hence the above measurement rules out the presence of magnetic clusters (along with glassy features) and brings out the atomic nature of spin arrangement in the sample; as the existence of clusters of spins will not allow total rejuvenation of the system. 

\subsection{Change in nature of magnetic ordering in the compound YMn$_{1.95}$Cr$_{0.05}$O$_5$}
Figure 3a shows the temperature response of $\chi$$_1$$^R$ of the Cr doped compound. 
A sharp step is observed around 56K which is followed by a peak at $\approx$ 33K, as the temperature is decreased. The ZFCM and FCM curves (at 200 Oe) of the compound as a function of temperature are shown in figure 3b. The features observed in low field ac susceptibility is completely changed in dc-magnetization and broad hump is observed in ZFCM around 27K, implying that higher magnetic field (even the field of the order of 200 Oe) suppresses the finer features in the sample. The bifurcation between the ZFCM and FCM curves starts $\approx$ 56 K, implying that this is the magnetic ordering temperature. This is also inferred from the d(M/H)/dT verses temperature plot which shows a peak around 56 K (not shown). Moreover, it is to be noted that absolute values of magnetization for YMn$_{1.95}$Cr$_{0.05}$O$_5$ is significantly higher than that of the parent compound. Hence it can be said that Cr substitution in YMn$_{2}$O$_5$, results in an increase in magnetic ordering temperature along with an increase in magnetization value. This fact is further substantiated from the MH isotherm at 10K, as shown in figure 3c. It is observed that with the Cr substitution the nature of curvature of the isotherm changes along with the increase in magnetization value. All these clearly indicate that the nature of magnetic ordering undergoes a major change with Cr substitution. Therefore, to discern the nature of magnetic ordering in YMn$_{1.95}$Cr$_{0.05}$O$_5$, temperature response of the real part of 2nd order of ac susceptibility ($\chi$$_2$$^R$) is studied. It is a well known fact that, $\chi$$_2$$^R$ arise due to the presence of a symmetry-breaking field and is only observed in the presence of external static magnetic field or in materials with permanent magnetization (internal field) like in ferromagnets or ferrimagnets and is absent in antiferromagnet or paramagnets. The $\chi$$_2$$^R$ is totally absent when measurements were done on YMn$_2$O$_5$, as expected because the compound undergoes an antiferromagnetic ordering. In contrast, a sharp peak is observed around 55K for YMn$_{1.95}$Cr$_{0.05}$O$_5$, when temperature response of the $\chi$$_2$$^R$ is noted (figure 3d). Presence of a significant $\chi$$_2$$^R$ signal near the magnetic transition temperature even in absence of external DC field supports the presence of symmetry breaking internal magnetic field. This symmetry breaking field appears from the spontaneous magnetization which is a consequence of uncompensated sublattice magnetization of ferrimagnetic system. In this context, it is to be noted that magnetization remains unsaturated even upto 100 kOe at 10K and it increases linearly with increasing field (inset fig 3c). This non saturation of magnetization at higher field is a typical character of ferrimagnetic system and is also observed in other ferrimagnets.\cite{bho} Hence from this measurement it can be said that substitution of Cr in YMn$_2$O$_5$ induces ferrimagnetic ordering in the compound. The hump observed at lower temperature $\approx$ 33K in $\chi$$_1$$^R$ might correspond to lock-in temperature associated with the ICM-CM transition, similar to the parent compound. In addition a peak in $\chi$$_2$$^R$ is observed around 35 K, possibly due to some symmetry breaking field arising out of Cr substitution. It can be said that this compound serves as a rare demonstration where there is a possibility of coexistence between ferrimagnetic ordering and ferroelectricity.

Further, similar to the parent compound, memory experiment is performed on YMn$_{1.95}$Cr$_{0.05}$O$_5$ with stopping at 35K and 25K (shown in Fig. 4). Aging effect is observed at both the waiting temperatures but unlike the parent compound the system does not rejuvenates to the zero age configuration when the field is switched on after the waiting time. During the warming cycle a distinct step like feature is seen only around 25K, while no such characteristic is observed around 35K. This implies that the system remembers its thermal history at 25K i.e. in the commensurate phase while in the incommensurate phase (i.e. at 35K), it is erased. The contrasting memory effect in the parent and the Cr substituted compound might arise possibly due to the presence of disorder in form Cr which introduces random interaction (both magnetic and electric) in the system. 

\section{Conclusion}In summary, low field magnetic studies of two compounds YMn$_2$O$_5$ and YMn$_{1.95}$Cr$_{0.05}$O$_5$ are carried out. It is shown that, in the parent compound, the evidences of multiple phase transitions involving changes in magnetic propagation vector are clearly reflected in low-field magnetic measurements. Through various time dependent magnetization protocols it is observed that the spin arrangements in the ICM and CM phase is different. Substitution of Cr at the Mn site changes the magnetic ordering of the parent compound from antiferromagnet to ferrimagnet with an enhanced ordering temperature. Direct evidence of appearance of spontaneous magnetization on Cr doping is given through observation of second order susceptibility. This observation in the Cr substituted compound arises due to magnetic interactions between Mn and Cr ions, without any significant change in structure. Moreover it give rise to memory effect of YMn$_{1.95}$Cr$_{0.05}$O$_5$ while such a feature is absent in the parent compound. Thus, this opens up a possible route to bring in the coexistence of ferroelectricity with ferromagnetism/ ferrimagnetism. 

\maketitle\section{Acknowledgement}
We are grateful to P. Chaddah for many fruitful discussions. We thank Puja Dey for introducing us to the pyrophoric method. KM acknowledges CSIR, India for fellowship.

\begin{figure*}
	\centering
		\includegraphics[width=8cm]{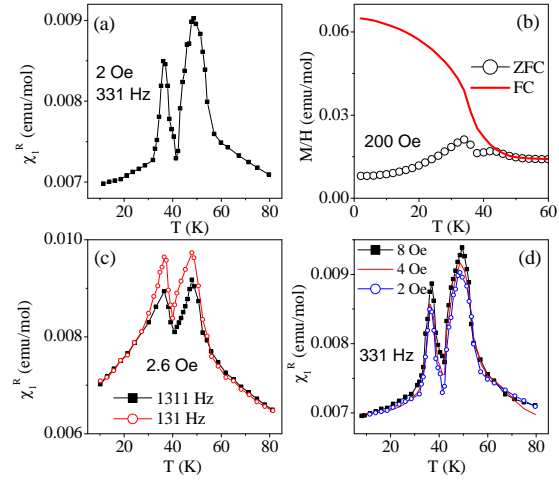}
	\caption{ For the YMn$_{2}$O$_5$ compound (a) Temperature response of real part of ac susceptibility ($\chi$$_1$$^R$). (b) Zero field cooled and Field cooled magnetization at 200 Oe static magnetic field. (c) and (d) Frequency and field dependence of ($\chi$$_1$$^R$) respectively. }
\end{figure*}
\begin{figure*}
	\centering
		\includegraphics[width=8cm]{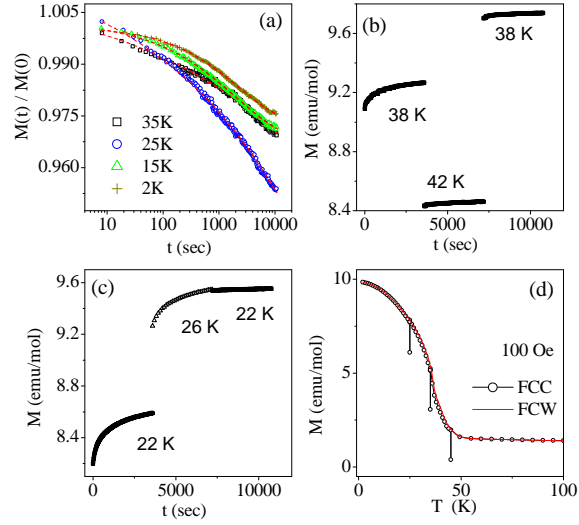}
	\caption{ (a) Time dependence of the magnetization for the compound YMn$_{2}$O$_5$ at different temperatures after cooling in a field of 10kOe. Dashed lines are fitting of the data of 35K and 25K to the equation (1) and while the data of 15K and 2K are fitted to the equation (2). (b) and (c) Magnetic relaxation in 500 Oe with temporary heating (t1, t2, t3 for one hour each) for ZFC case in incommensurate phase and commensurate phase respectively. (d) Magnetization verses temperature curves during field cooling. The field is switched off at three temperatures (45K, 35K and 25K) for a waiting time of 7200 s. Magnetization verses temperature curve in warming mode is also shown which is seen to totally superimpose on the cooling curve}
\end{figure*}
\begin{figure*}
	\centering
		\includegraphics[width=8cm]{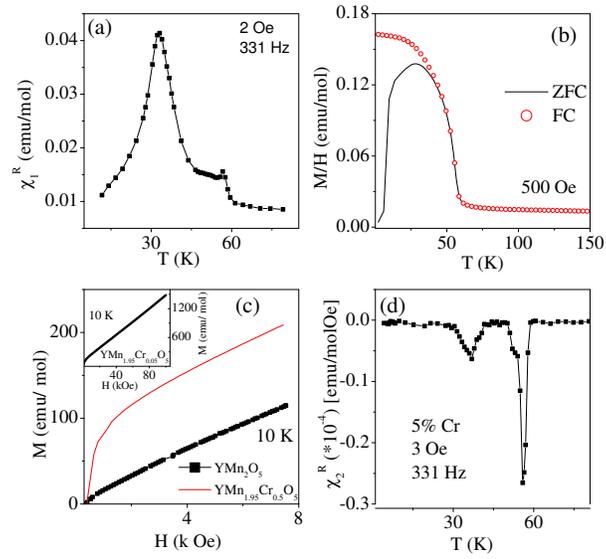}
	\caption{(a)Temperature response of real part ($\chi$$_1$$^R$) of ac susceptibility for the compound YMn$_{1.95}$Cr$_{0.05}$O$_5$. (b) Zero field cooled and Field cooled magnetization for the same sample at 500 Oe. (c) Magnetization as a function of magnetic field at 10K for both compounds YMn$_{2}$O$_5$ and YMn$_{1.95}$Cr$_{0.05}$O$_5$. Inset: Magnetization as a function of magnetic field at 10K for YMn$_{1.95}$Cr$_{0.05}$O$_5$ upto 100 kOe. (d)Temperature response of real part of second order of ac susceptibility ($\chi$$_2$$^R$) for the compositions YMn$_{1.95}$Cr$_{0.05}$O$_5$.}
\end{figure*}
\begin{figure*}
	\centering
		\includegraphics[width=8cm]{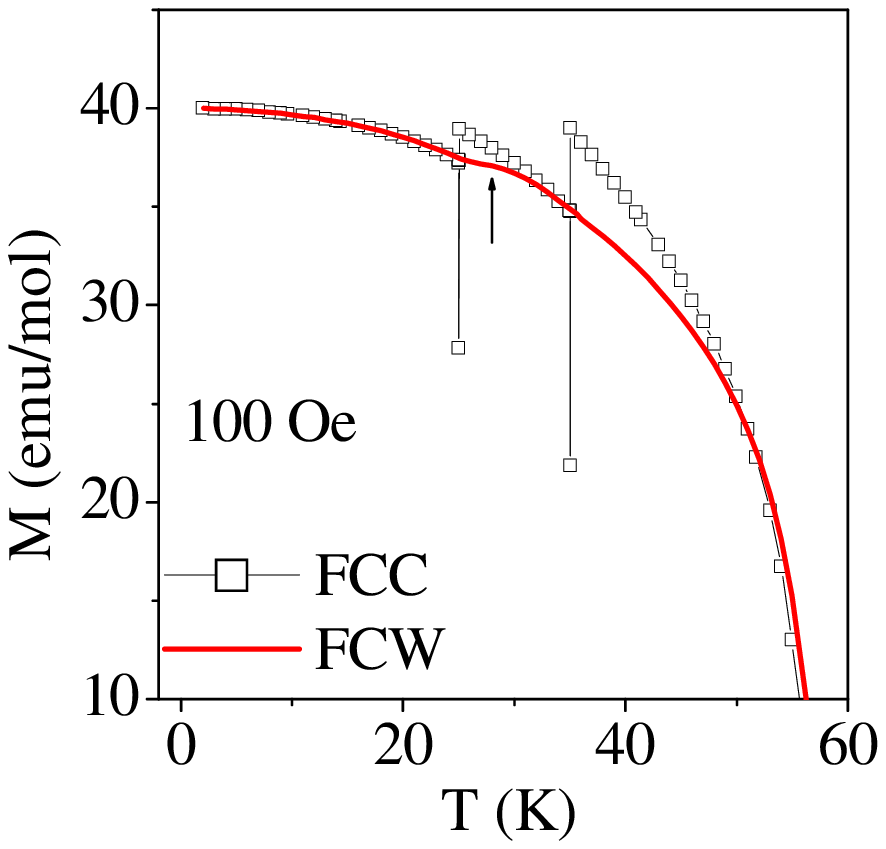}
	\caption{ Magnetization as a function of temperature during field cooling for the sample YMn$_{1.95}$Cr$_{0.05}$O$_5$ (symbols). The field is switched off at two temperatures (35K and 25K) for a waiting time of 7200 s. The field cooled warming curve is shown by the continuous line. The arrow points to the kink in the warming curve indicating the memory effect.}
\end{figure*}

\begin{thebibliography}{}
\bibitem[1]{lee} J. H. Lee, L. Fang, E. Vlahos, X. Ke, Y. W. Jung, L. F. Kourkoutis, J-W Kim, P. J. Ryan, T. Heeg, M. Roeckerath, V. Goian, M.Bernhagen, R. Uecker, P. C. Hammel, K. M. Rabe, S. Kamba, J. Schubert, J. W. Freeland, D. A. Muller, C. J. Fennie, P. Schiffer, V. Gopalan, E. J. Halperin, and D. G. Schlom, Nature (London) \textbf {466}, 954 (2010); R. Ramesh, Nature (London) \textbf {461}, 1218 (2009), B.van Aken, T. T. M. Palstra, A. Fillippetti, and N. A. Spaldin, Nature Mater \textbf {3}, 164 (2004)
\bibitem[2]{kim} T. Kimura T. Goto, H. Shintani, K. Ishizaka, T. Arima, and Y. Tokura, Nature (London) \textbf {426}, 55 (2003)
\bibitem[3]{hur} N.Hur, S. Park, P. A. Sharma, J. S. Ahn, S. Guha, and S-W. Cheong, Nature (London) \textbf {429}, 392 (2004)
\bibitem[4]{cha} L.C. Chapon, G. R. Blake, M. J. Gutmann, S. Park, N. Hur, P.G. Radaelli, and S-W. Cheong, Phys. Rev. Lett. \textbf {93}, 177402 (2004)
\bibitem[5]{bla} G.R. Blake, L. C. Chapon, P. G. Radaelli, S. Park, N. Hur, S-W. Cheong, and J. Rodríguez-Carvajal, Phys. Rev. B.\textbf {71}, 214402 (2005)
\bibitem[6]{cha1} L. C. Chapon, P. G. Radaelli, G. R. Blake, S. Park, and S.-W. Cheong, Phys. Rev. Lett. \textbf {96}, 097601 (2006)
\bibitem[7]{yam} Y. Yamasaki, S. Miyasaka, Y. Kaneko, J.-P. He, T. Arima, and Y. Tokura, Phys. Rev. Lett. \textbf {96}, 207204 (2006)
\bibitem[8]{sus} A. B. Sushkov, R. V. Aguilar, S. Park, S-W. Cheong, and H. D. Drew, Phys. Rev. Lett. \textbf {98}, 027202 (2007)
\bibitem[9]{nod} Y. Noda, H. Kimura, M. Fukunaga, S. Kobayashi, I. Kagomiya and K. Kohn J. Phys.:  Condens. Matter \textbf {20}, 434206 (2008)
\bibitem[10]{woo} W. S. Choi, S. J. Moon, S. S. A. Seo, D. Lee, J. H. Lee, P. Murugave, T. W. Noh, and Y. S. Lee, Phys. Rev. B. \textbf {78}, 054440 (2008)
\bibitem[11] {fab} X. Fabreges, S. Petit, I. Mirebeau, S. Pailhes, L. Pinsard, A. Forget, M. T. Fernandez-Diaz, and F. Porcher, Rev. Lett. \textbf {103}, 067204 (2009) 
\bibitem[12]{mam} M. Fukunaga and Y. Noda, J. Phys. Soc. Jpn. \textbf  {79}, 054705 (2010)
\bibitem[13]{kho} D. I. Khomskii J. Magn. Magn. Mater \textbf {306}, 001 (2006)
\bibitem[14]{ino} A. Inomata and and K. Kohn, J. Phys.:  Condens. Matter \textbf {8}, 2673 (1996)
\bibitem[15]{hur1} N. Hur, S. Park, P. A. Sharma, S. Guha, and S-W. Cheong, Phys. Rev. Lett. \textbf {93}, 107207 (2004) 
\bibitem[16]{kim1} H. Kimura, Y. Kamada, Y. Noda, K. Kaneko, N. Metoki and K. Kohn, J. Phys. Soc. Jpn. \textbf  {75}, 113701 (2006)
\bibitem[17]{hiro} H. Kimura, S. Kobayashi, Y. Fukuda, T. Osawa, Y. Kamada, Y. Noda, I. Kagomiya, and K. Kohn, J. Phys. Soc. Jpn. \textbf  {76}, 074706 (2007)
\bibitem[18]{sou} R. A. de Souza, U. Staub, V. Scagnoli, M. Garganourakis, Y. Bodenthin, S.-W. Huang, M. Garcýa-Fernandez,S. Ji, S.-H. Lee, S. Park, and S.-W. Cheon, Phys. Rev. B. \textbf {84}, 104416 (2011)
\bibitem[19]{par} S. Partzsch, S. B. Wilkins, J. P. Hill, E. Schierle, E. Weschke, D. Souptel, B. Buchner, and J. Geck, Phys. Rev. Lett. \textbf {107}, 057201 (2011)
\bibitem[20]{kim3} J.-H. Kim, M. A. van der Vegte, A. Scaramucci, S. Artyukhin, J.-H. Chung, S. Park, S-W. Cheong,M. Mostovoy, and S.-H. Lee, Phys. Rev. Lett. \textbf {107}, 097401 (2011) 
\bibitem[21]{kim2} T. Kimura, Y. Tomioka, R. Kumai, Y. Okimoto, and Y. Tokura, Phys. Rev. Lett. \textbf {83}, 3940 (1999), C.N. R. Rao and B. Raveau, Colossal Magnetoresistance, Charge Ordering and Related Properties of Manganese Oxides, World Scientific, Singapore (1998).
\bibitem[22]{zhou} H. D. Zhou, J. C. Denyszyn, and J. B. Goodenough, Phys. Rev. B. \textbf {72}, 224401 (2005), C. de la Calle, J.A. Alonso, M.J. Martínez-Lope, M. García-Hernández and G. André, Materials Research Bulletin \textbf {43} 197 (2008), F. Wunderlich, T. Leisegang, T. Weißbach, M. Zschornak, H. Stöcker, J. Dshemuchadse, A. Lubk, T. Führlich, E. Welter, D. Souptel, S. Gemming, G. Seifert, and D. C. Meyer, Phys. Rev. B. \textbf {82}, 014409 (2010)
\bibitem[23]{dey} R. K. Patil, J. C. Ray and P. Pramanik J. Am. Ceram. Soc. \textbf {84}, 2849 (2001). 
\bibitem[24]{you}	R. A. Young, A. Sakthivel, T.S. Moss, C.O. Paiva-Santos, User Guide To Program DBWS-9411, Georgia Institute of technology, Atlanta, (1994)
\bibitem[25]{ashnarsi} A. Bajpai and A. Banerjee, Rev. Sci. Instrum. \textbf{68}, 4075 (1997).
\bibitem[26]{sat} S. Kobayashi, T. Osawa, H. Kimura, Y. Noda, I. Kagomiya and K. Kohn J. Phys. Soc. Jpn. \textbf  {73}, 1593 (2004)
\bibitem[27]{akm} G. Sinha, R. Chatterjee, M.Uehara and A. K. Majumder J. Magn. Magn. Mater \textbf {164}, 345 (1996)
\bibitem[28]{muk} K. Mukherjee and A. Banerjee Phys. Rev. B. \textbf {77}, 024430 (2008) and references therein.
\bibitem[29]{bho} R. N. Bhowmik and R. Ranganathan, Phys. Rev. B. \textbf {74}, 214417 (2006).



\end{thebibliography}
\end{document}